\documentclass[a4paper]{article}
\usepackage{graphicx}
\usepackage{xspace}

\newcommand{\FlukaLong}{{\scshape Fluka2008}\xspace}
\newcommand{\UrqmdLong}{{\scshape Urqmd1.3.1}\xspace}
\newcommand{\VenusLong}{{\scshape Venus4.12}\xspace}

\begin{document}

\begin{center}
{\Large \bf Tuning of the GEANT4 FRITIOF (FTF) Model Using NA61/SHINE Experimental Data}
\end{center}

\begin{center}
{V. Uzhinsky}
\end{center}

\begin{center}
{CERN, Geneva, Switzerland and LIT, JINR, Dubna, Russia}
\end{center}

\begin{center}
(on behalf of Geant4 hadronic group)
\end{center}

\begin{center}
\begin{minipage}{12cm}
The NA61/SHINE collaboration measured inclusive cross sections of $\pi^+$ and $\pi^-$
meson production in the interactions of 31 GeV/c protons with carbon nuclei at forward
emission angles (0 -- 420 {\it mrad}). The collaboration also presented predictions
of Monte Carlo models -- FLUKA, VENUS and UrQMD, in comparison with the data.
A careful analysis shows that deviations of the FLUKA and VENUS predictions from the data have
different tendencies. The worst description of the data was observed for the UrQMD model
results.

~~~~~~~~All the models assume the creation of quark-gluon strings in the interactions,
but it is complicated to analyze the models in order to find the source of the deviations.
Thus, the quark-gluon string model -- FRITIOF (FTF) -- was implemented in the GEANT4 toolkit
and is used to understand the deviations mentioned above.  It was found that the most
important factor influencing the FTF calculations is the sampling of quark-gluon string
masses. The other factors/parameters are not essential for a description of the data.
Also, a good description of the data is achieved by the FTF model.

\end{minipage}
\end{center}

Some selected results of the NA61/SHINE collaboration \cite{NA61} in comparison with the
model predictions obtained by the collaboration, and taken from Fig. 23 of Ref. \cite{NA61}
together with its figure caption, are reproduced in Fig. 1.
\begin{figure}[cbth]
\includegraphics[width=160mm,height=90mm,clip]{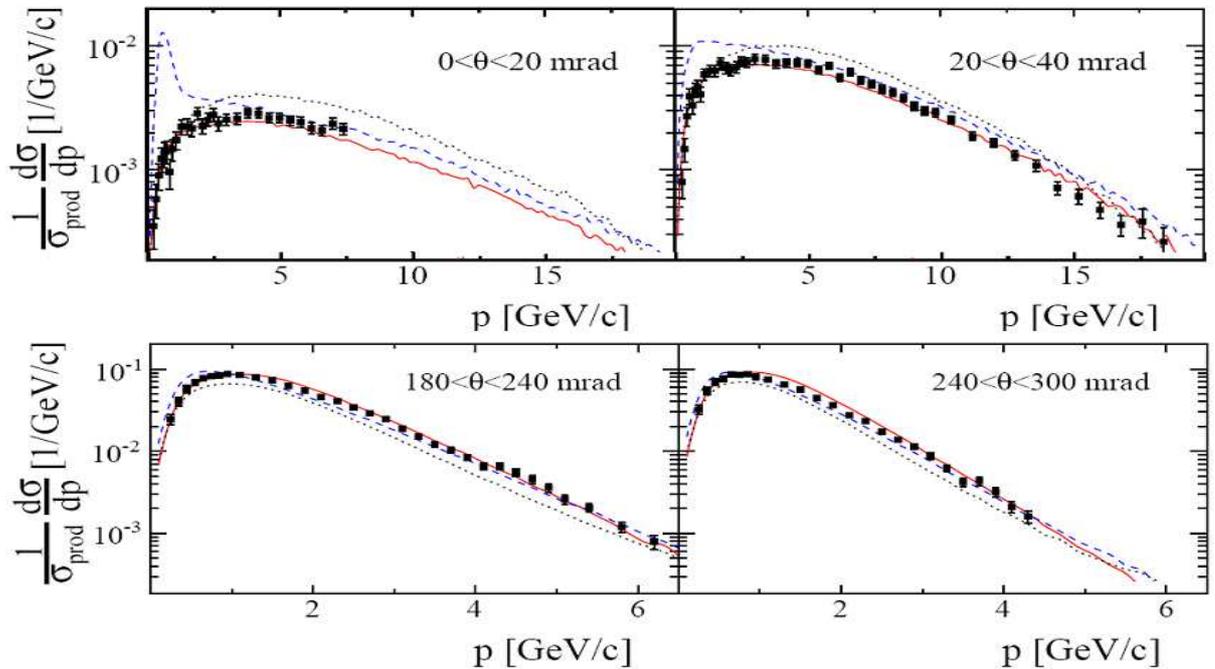}
\caption{
Laboratory momentum distributions of $\pi^{+}$ mesons produced in production p+C interactions at
31~GeV/c in different intervals of polar angle ($\theta$).
The spectra are normalized to the mean $\pi^{+}$ multiplicity in
all production p+C interactions. Error bars indicate statistical and systematic uncertainties
added in quadrature.  The overall uncertainty~($2.3\%$) due to the normalization procedure
is not shown.  Predictions of hadron production models, \FlukaLong (solid line),
\UrqmdLong (dashed line) and \VenusLong (dotted line) are also indicated.
}
\label{Fig1}
\end{figure}

It is difficult to estimate the size of the model prediction deviations from the data given
the choice of scale and normalization.  Though, it can be seen that at small angles
the VENUS model \cite{Venus} overestimates the data.  At large angles, the model predictions
underestimate the data.  So, the meson yield as a function of angle decreases faster in
the model than in the experimental data.

The FLUKA model \cite{Fluka} underestimates the data at small angles a little bit. At the
same time, the model overestimates the data at large angles. So, the meson yield versus 
angle decreases more slowly in the model than in the experimental data.

The UrQMD model \cite{UrQMD} predictions are not considered in the present work.  A special
paper \cite{Uzhi} is devoted to this subject.

The FTF model predictions with absolute normalization are shown in Fig. 2.
\begin{figure}[cbth]
\includegraphics[width=160mm,height=100mm,clip]{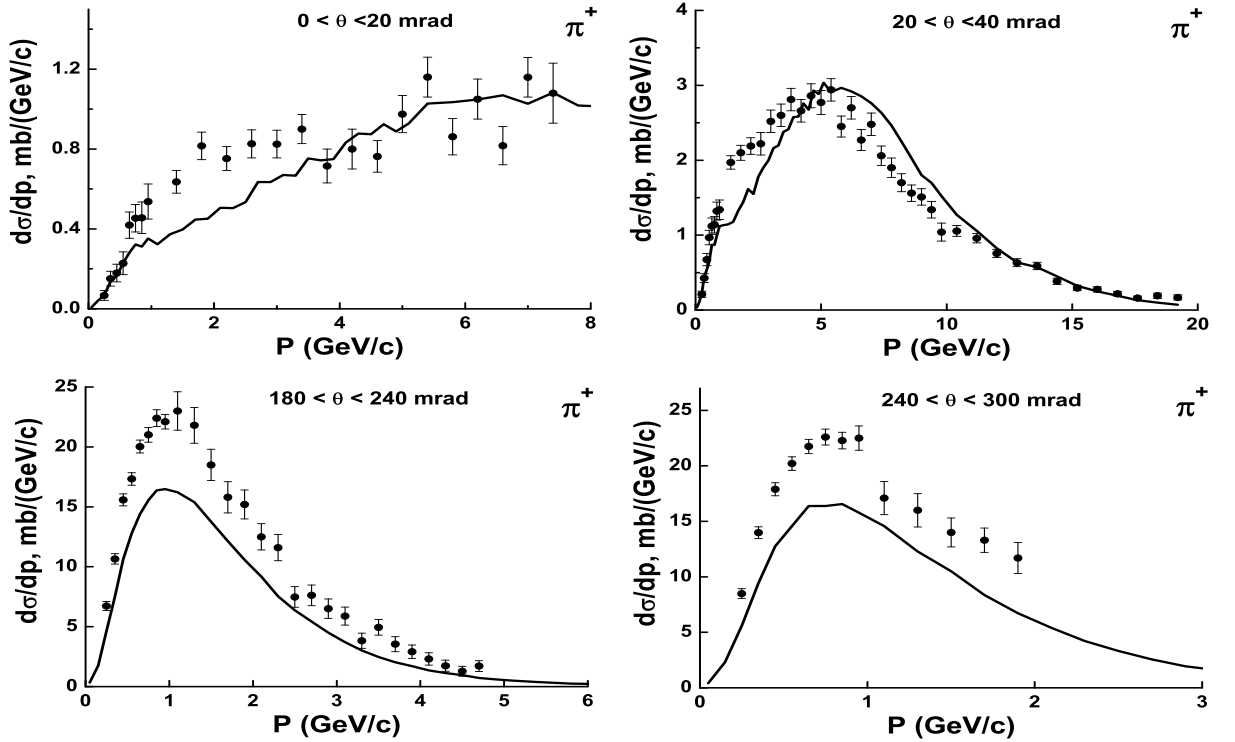}
\caption{Inclusive cross sections of $\pi^+$ meson production.
Points are experimental data \protect{\cite{NA61}}. Lines are the FTF model calculations.}
\label{Fig2}
\end{figure}
As seen, the model is in an agreement with the data at small angles. At large angles, the model
underestimates the data. Thus, FTF has the same tendency as VENUS.  A subject of
the present paper is an improvement of the FTF model.

\begin{figure}[cbth]
\begin{center}
\includegraphics[width=100mm,height=30mm,clip]{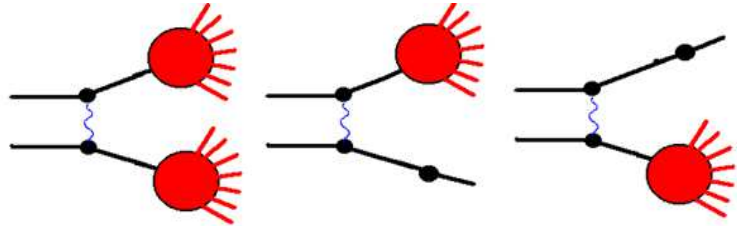}
\caption{Double diffraction and single diffraction considered in the Fritiof model.}
\label{Fig3}
\end{center}
\end{figure}
The FTF model is based on the well-known FRITIOF model \cite{Fritiof} (see also Ref. 
\cite{FTFcode}). The Fritiof model assumes that all hadron-hadron interactions are binary
reactions, $h_1+h_2 \rightarrow h_1'+h_2'$, where $h_1'$ and $h_2'$ are excited states of 
the hadrons with continuous mass spectra (see Fig. 3). If one of the post-interaction 
hadrons remains in the ground state ($h1 + h2 -> h1 + h2'$) the reaction is referred to 
as "single diffraction dissociation".  If neither hadron remains in the ground state 
the reaction is referred to as "double diffraction dissociation". The excited hadrons 
are considered as QCD-strings, and the corresponding LUND-string fragmentation model is 
applied for a simulation of their decays.

The key ingredient of the Fritiof model is the sampling of the string masses. In general, 
the set of possible final states of the interaction can be represented by Fig. 4, 
where samples of possible string masses are shown. There is a point corresponding to 
elastic scattering, a group of points which represents final states of binary 
hadron-hadron interactions ($h_1+h_2 \rightarrow h_3+h_4$), lines corresponding to
the single diffractive interactions, and various intermediate regions. The region populated
with the brown points is responsible for the double diffractive interactions.  
In principle the mass sampling thresholds may be below the initial hadron masses.
The string masses are sampled in the triangular region restricted by the diagonal line
corresponding to the kinematical limit $M_1+M_2 =E_{cms}$, and the threshold lines.
If a point is below string mass threshold, it is shifted to a nearest diffraction line.
\begin{figure}[cbth]
\begin{center}
\includegraphics[width=100mm,height=80mm,clip]{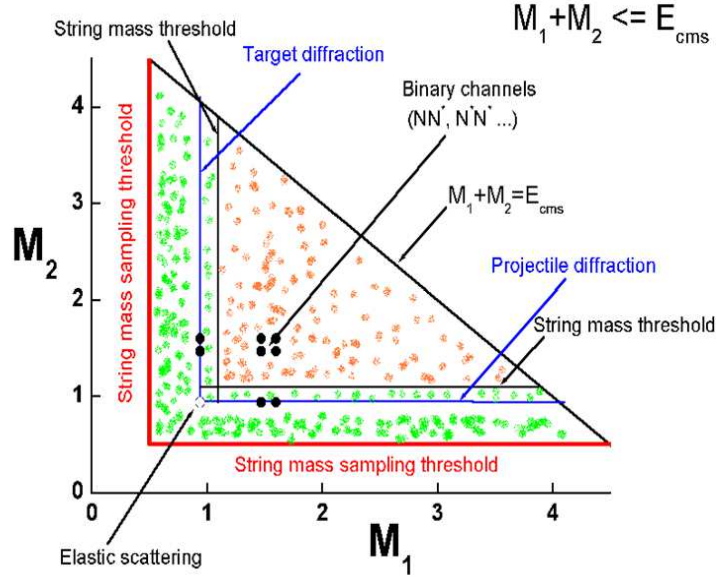}
\caption{Diagram of final states of nucleon-nucleon interactions. The axes are in GeV.}
\label{Fig4}
\end{center}
\end{figure}

In the Fritiof model code implemented in Geant4 we take into account binary 
reaction final states differently from elastic scattering. Weights of the hN elastic
scattering and the hN diffractive interactions have been introduced which give probabilities 
of the corresponding reactions. In the case of single diffraction we sample the points of 
the diagrams only on the diffraction lines.

The probability of the single diffraction ($W_{sd}$) at high energies is given as:
$$W_{sd}=D\ exp(-d\ Y_{lab}),\ \ \ D=0.805,\ \ \ d=0.35,$$
where  $Y_{lab}$ is the projectile rapidity in the target rest frame.
The squared mass of the string created in the reaction is equal to
$$(M)^2=P\ \left[E_{CMS} - \frac{m_0^2 + P_T^2}{E_{CMS} - P}\right] - P_T^2,$$
where $P$ is the light-cone momentum sampled according to the distribution $dP/P$.
$P=E-P_z$ for the hadron originating from the projectile, and $P=E+P_z$ for the 
hadron originating from the target, in the CMS. 
$m_0$ is the mass of the hadron left in the ground state. $P_T^2$ is the square of the
transferred transverse momentum, $<P_T^2>=$0.15 $(GeV/c)^2$.

The probability of double diffraction, $W_{dd}=1-2 W_{sd}$, for nucleon-nucleon
interactions.
In double diffraction $P^-$, the momentum of the hadron originating from the projectile
and $P^+$, the momentum of the hadron originating from the target, are sampled independently
according to the distribution $dW \propto dP/P$.  The string mass sampling threshold,
 $m_{th}=m_0+280$ {\it MeV}.

The Fritiof model assumes that in the course of a hadron-nucleus interaction the string
originating from the projectile can interact with various intra-nuclear nucleons and 
thus reach highly excited states. The probability of multiple interactions is 
calculated in the simplest approximation. The cascading of secondary particles is neglected 
as a rule.  Due to these assumptions, the original Fritiof model fails to describe nuclear
destruction and slow particle spectra. In order to overcome these difficulties we enlarge
the model by including the reggeon theory-inspired model of nuclear
desctruction \cite{RTIM1,RTIM2}. Momenta of the nucleons ejected from a nucleus during the
"reggeon cascade" are sampled according to a "Fermi motion" algorithm presented in
Ref. \cite{FermiM}.

The main parameters of the model are the transferred transverse momentum and the probability
of single diffraction at a given energy.  The form of the light-cone momentum distribution
can be changed also.

It was checked that decreasing (increasing) $<P_T^2>$ leads to an increase (decrease) of
the inclusive cross section maxima at $\theta < $ 140 {\it mrad}. 
There is no effect at larger angles.

Variations of $W_{sd}$ change the integrals of the cross sections without changing
their forms.

The last possibility -- a change of the form of the light-cone momentum distribution, turns
out to be successful.  Calculations performed assuming a flat P distribution ($dW \propto dP$) 
for the double diffraction dissociation processes are shown in Fig. 5 by solid lines. As seen,
they are rather close to the experimental data.
\begin{figure}[cbth]
\includegraphics[width=160mm,height=100mm,clip]{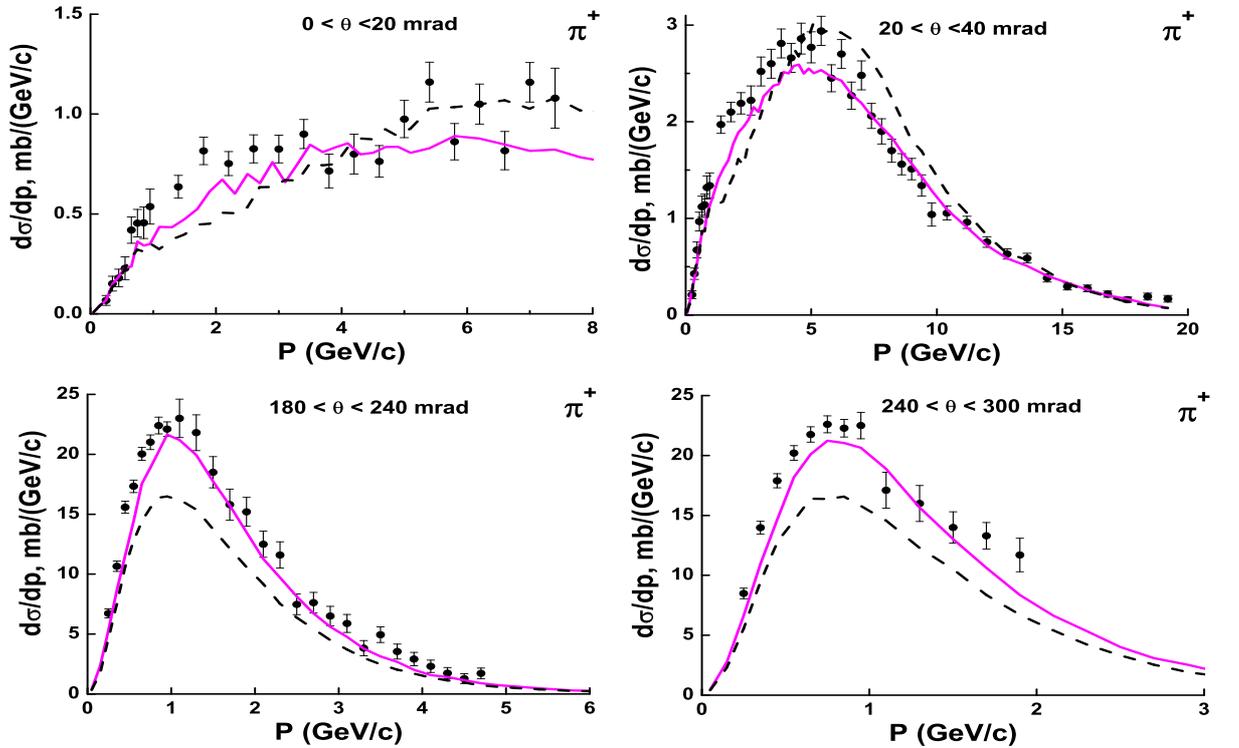}
\caption{Inclusive cross sections of $\pi^+$ meson production.
Points are experimental data \protect{\cite{NA61}}. Solid and dashed lines are the FTF model
calculations with flat and $1/P$ distributions, respectively.}
\label{Fig5}
\end{figure}

It is obvious that average masses of strings are larger with the flat distribution than with 
the $1/P$ distribution. Thus, multiplicities of produced particles in the target fragmentation
region, at large angles and small momenta, are larger in the first case. At the same time, 
the multiplicities in the projectile fragmentation region, at small angles and large momenta,
become lower due to kinematics with the flat distribution.

From this point of view, we can suppose that the masses of the strings are larger than 
needed for a description of the data in the FLUKA model. Results of the VENUS model point 
to low string masses in the model.  All of the features of the model can be improved.

Note that until now there were only restricted possibilities to check string mass
distributions. In this sense, the NA61/SHINE data are unique. It would be well to 
continue the experimental studies, reach higher statistics, and extend the measurements on 
K-mesons and protons. The proton spectra are extremely interesting because Monte Carlo models
give very different predictions for them.

\section*{Conclusion}
\begin{enumerate}
\item To reach agreement of the Geant4 FTF model with the experimental data of the NA61/SHINE
collaboration, it is necessary to replace the $1/P$ distribution of the squared string masses
by the flat distribution in the double diffraction dissociation processes.

\item The string masses sampled in the FLUKA model are probably a little bit higher than 
necessary for a correct description of the data.

\item The string masses sampled in the VENUS model are too small.

\end{enumerate}

The author is thankful to D.H. Wright, G. Folger and V. Ivanchenko for interest in this work 
and important remarks.

\end{document}